\documentclass[twocolumn,showpacs,preprintnumbers,amsmath,amssymb]{revtex4}

\usepackage{graphicx}
\usepackage{dcolumn}
\usepackage{bm}

\begin{document}

\draft

\title{Manipulation of a single charge in a double quantum dot}
\author{J.~R.~Petta}
\author{A.~C.~Johnson}
\author{C.~M.~Marcus}
\affiliation{Department of Physics, Harvard University, Cambridge,
MA 02138}
\author{M.~P.~Hanson}
\author{A.~C.~Gossard}
\affiliation{Materials Department, University of California, Santa
Barbara, California 93106}

\date{\today}

\begin{abstract}
We manipulate a single electron in a fully tunable double quantum
dot using microwave excitation. Under resonant conditions,
microwaves drive transitions between the (1,0) and (0,1) charge
states of the double dot. Local quantum point contact charge
detectors enable a direct measurement of the photon-induced change
in occupancy of the charge states. From charge sensing
measurements, we find $T_1$$\sim$16 ns and a lower bound estimate
for $T{_2^*}$ of 400 ps for the charge two-level system.

\end{abstract}


\maketitle Mesoscopic circuits can be designed to create
artificial two-level systems that can be controlled on nanosecond
time scales, allowing the observation of coherent oscillations
between the two quantum states
\cite{Devoret_Science_2002,Nakamura_Nature_1999}. A broad range of
experiments have demonstrated control over the flux states of a
SQUID \cite{Van_der_Wal_Science_2000}, the phase of a
Josephson-junction qubit \cite{Yu_Science_2002,Martinis_PRL_2002},
and the charge states of semiconducting quantum dots
\cite{Hayashi_PRL_2003}.

Semiconducting quantum dots are promising systems for the
manipulation of a single charge because of the relative ease of
controlled confinement using electrostatic gates
\cite{Ciorga_PRB_2000,Kouwenhoven_RPP_2001}. By this same
approach, coupled quantum dots can be used to create two-level
systems with precise and rapid control of the coupling between
quantum states \cite{Van_der_Wiel_RMP_2003}. Careful control of
the interdot tunnel coupling makes it possible to tune from a
weakly coupled regime, where a single charge is localized on one
of the dots, to a strongly coupled regime, where the charge
becomes delocalized \cite{Oosterkamp_Nature_1998}. Under resonant
conditions, microwaves can induce transitions between the charge
states, resulting in controlled charge state repopulation
\cite{Tien_PR_1963,Kouwenhoven_PRL_1994,Fujisawa_SuperlatMicro_1997}.

In this Letter, we create a two-level system from a double quantum
dot containing a single electron. We drive resonant transitions
between the charge states through the application of microwaves.
Previously, photon assisted tunneling (PAT) has been used to
detect microwave excitation of a many-electron double quantum dot
\cite{Oosterkamp_Nature_1998}. Here we directly measure the
occupancy of the charge states using local quantum point contact
(QPC) charge detectors \cite{Field_PRL_1993}. From these
measurements we have extracted the lifetimes $T_1$ and $T_{2}^{*}$
for a semiconductor dot based charge two-level system
\cite{Lehnert_PRL_2003}. In contrast with PAT, which requires
coupling to the leads, our sensing technique can be used in
regimes where transport is not possible, allowing for spectroscopy
of an isolated double dot.

Measurements are performed on gate-defined quantum dots fabricated
on a GaAs/Al$_{0.3}$Ga$_{0.7}$As heterostructure grown by
molecular beam epitaxy (Fig.\ 1(a)). A two-dimensional electron
gas with electron density 2$\times$10$^{11}$cm$^{-2}$ and mobility
2$\times$10$^5$cm$^2$/V$\cdot$s lies 100 nm below the surface and
is depleted with Ti/Au top gates. Gates 2--6 and $t$ form the
double quantum dot. Gates 3--5 are connected via bias tees to dc
and microwave sources \cite{Anritsu}. QPC charge detectors are
created by depleting gates 1 and 7, while gate 8 is energized to
isolate the QPC sensor from the double dot circuitry. Gates 9--11
are unused.

\begin{figure}[b]
\vspace{1.6 cm}
\begin{center}
\includegraphics[width=8.6cm]{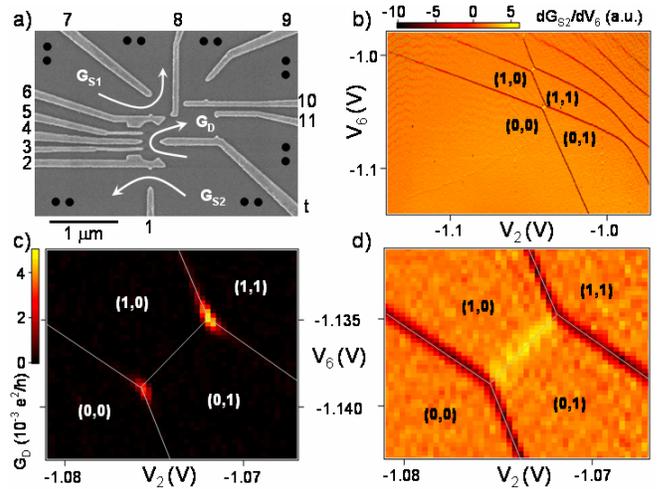}
\end{center}
\vspace{-2.5 cm} \caption{(a) SEM image of a device identical in
design to the one used in this experiment. Gates 2--6 and $t$
define the double dot. QPC charge detectors are formed by
depleting gates 1 and 7. Gate 8 may be used to separate the QPC
and double dot conductance measurement circuits. Gates 9--11 are
not energized. $\bullet$ denotes an ohmic contact. (b) Large scale
plot of $dG_{S2}$/$dV_{6}$ as a function of $V_{2}$ and $V_{6}$.
Charge states are labelled $($$M$,$N$$)$, where $M$$($$N$$)$ is
the time averaged number of electrons on the upper (lower) dot.
$G{_D}$, in (c), and $dG_{S2}$/$dV_{6}$, in (d), as a function of
$V_{2}$ and $V_{6}$ near the (1,0) to (0,1) transition. In (c--d)
the gates have been slightly adjusted relative to (b) to allow
simultaneous transport and sensing. Identical color-scales are
used in (b) and (d).}\vspace{-0.5 cm}
\end{figure}

The double dot conductance, $G_{D}$, and the QPC conductances,
$G_{S1(S2)}$, are measured using standard ac lock-in techniques
with the sample cooled to base temperature in a dilution
refrigerator. The electron temperature, $T_e$$\sim$135 mK, was
determined from Coulomb blockade peak widths.  The double dot is
voltage biased with a 6 $\mu$V excitation at 17 Hz, while the QPC
detectors are current biased at 1 nA at frequencies of 93 and 97
Hz. This setup allows a simultaneous measurement of $G_{D}$,
$G_{S1}$, and $G_{S2}$.

Transport in the few-electron regime is made difficult by the
reduction in tunnel coupling to the leads as the dot is depleted
\cite{Ciorga_PRB_2000}. However, recent experiments using both
charge sensing and transport have shown that it is possible to
create a few electron double dot without sacrificing transport
\cite{Elzerman_PRB_2003}. We demonstrate similar control in Fig.\
1 (b--d). Figure 1(b) shows $dG_{S2}$/$dV_6$ (numerically
differentiated) as a function of $V_2$ and $V_6$. Electrons
entering or leaving the double dot, or moving from one dot to the
other, change the QPC conductance. These changes show up clearly
in the gate voltage derivatives of $G_{S1}$ and $G_{S2}$, and
directly map out the charge stability diagram of the double dot.
The nearly vertical lines correspond to charge transitions in the
lower dot, while the nearly horizontal lines are due to charge
transitions in the upper dot. In the lower left corner of the
charge stability diagram, the double dot is completely empty,
denoted (0,0). With the device configured as in Fig.\ 1(b), the
transport signal near the (1,0) to (0,1) transition is below the
noise floor of the measurement. A slight retuning of the gates
results in transport. Figure 1(c) shows a color scale plot of
$G_D$ near the (1,0) to (0,1) charge transition. A simultaneously
acquired charge stability diagram is shown in Fig.\ 1(d). In the
remainder of the paper, we will focus on the (1,0) to (0,1) charge
transition. Crossing this transition by making $V_6$ more positive
transfers a single electron from the lower dot to the upper dot.
This increases $G_{S2}$, resulting in the yellow line in the
charge stability diagram. In contrast, the dark lines correspond
to charge transitions that increase the total number of electrons
on the double dot as $V_6$ is increased, resulting in a decrease
in $G_{S2}$.

Near the interdot transition, the double dot forms a two-level
charge system that can be characterized by the detuning parameter,
$\epsilon$, and the tunnel coupling, $t$ (see inset of Fig. 3(d))
\cite{Van_der_Wiel_RMP_2003}. Tuning $t$ controls the crossover
from localized to delocalized charge states
\cite{Oosterkamp_Nature_1998}. This tunability is important,
because proposals involving the manipulation of electron spin in a
double dot often require control of the exchange interaction,
$J$=$4t^2/U$ \cite{Loss_PRA_1998}. We demonstrate control of $t$
in the one-electron regime in Fig.\ 2. As $V_t$ is increased, the
interdot charge transition smears out due to the onset of charge
delocalization (compare the upper and lower insets). A
quantitative measure of $t$ is made by measuring the QPC response
along the detuning diagonal (a typical detuning sweep is indicated
by the black line in the lower inset of Fig.\ 2). The QPC response
is converted into units of charge following DiCarlo \textit{et
al.} \cite{DiCarlo_PRL_2004}. Figure 2 shows $M$ as a function of
$\epsilon$ for several values of $V_t$. We fit the experimental
data using \cite{DiCarlo_PRL_2004}:
\begin{equation}
M=\frac{1}{2}\left(1-\frac{\epsilon}{\sqrt{\epsilon^2+4t^2}} \tanh
\left(\frac{\sqrt{\epsilon^2+4t^2}}{2k_B T_e}\right)\right)
\end{equation}
where $t$ is a free parameter and $k_B$ is Boltzman's constant.
$\epsilon$ is converted into units of energy by multiplying by the
lever arm (which takes into account capacitive division of
voltage). With $V_t$=-1.08 V, the QPC response is temperature
broadened, which places an upper limit on 2$t$ of 5.6 GHz. A 30 mV
increase in $V_t$ more than doubles $t$ (see the table in Fig.\
2).

\begin{figure}[t]
\vspace{1.55 cm}
\begin{center}
\includegraphics[width=8.6cm]{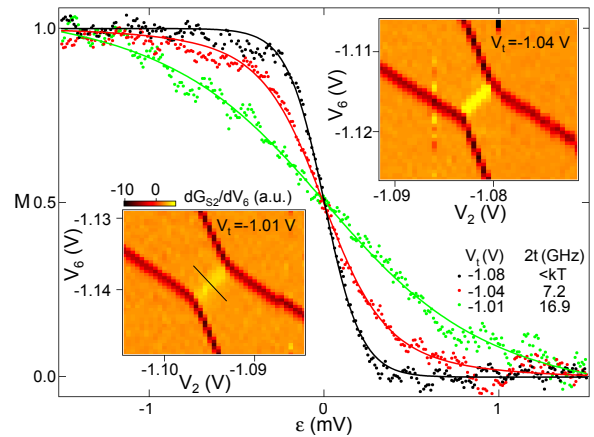}
\end{center}
\vspace{-2.8cm} \caption{ The number of electrons on the upper
dot, $M$, as a function of detuning, $\epsilon$, for several
values of $V_t$. Solid lines are best fits to the data (see text).
Increasing $V_t$ strengthens the interdot tunnel coupling and
broadens the interdot charge transition. A typical detuning sweep
follows the black line in the lower inset. Insets: Plots of
$dG_{S2}$/$dV_{6}$ as a function of $V_{2}$ and $V_{6}$ for
$V_t$=-1.01 V (lower inset) and $V_t$=-1.04 V (upper inset). The
same color-scale is used in both insets.} \vspace{-0.5cm}
\end{figure}

Microwaves can induce a current at zero source-drain bias when the
photon frequency is equal to the energy separation between the
(1,0) and (0,1) charge states
\cite{Stafford_PRL_1996,Stoof_PRB_1996,Brune_PhysicaE_1997}. This
PAT current is shown in Fig.\ 3(a) as a function of $V_2$ and
$V_6$ with 24 GHz continuous-wave (cw) applied to gate 4 of the
device and $V_{sd}$=0 $\mu$V. Four sharp resonances appear in the
vicinity of each triple point. The resonances closest to each
triple point correspond to 1 photon (1$\gamma$) processes, while
the outer resonances correspond to 2$\gamma$ processes, consistent
with previous observations of PAT in a semiconductor double dot
\cite{Oosterkamp_Nature_1998,Elzerman_PRB_2003}.

Charge sensing allows a direct measurement of microwave-induced
charge state repopulation
\cite{Van_der_Wal_Science_2000,Lehnert_PRL_2003}. Fig.\ 3(b) shows
a color-scale plot of $\delta$$G_{S1}$ as a function of $V_2$ and
$V_6$ with 24 GHz cw applied to gate 4 of the device and
$V_{sd}$=0 $\mu$V. Four stripes, aligned parallel to the (1,0) to
(0,1) charge transition line, appear in the presence of
microwaves. These features are absent when no microwave power is
applied (Fig.\ 1(d)). We associate these features with 1$\gamma$
and 2$\gamma$ processes that drive an electron from the (1,0)
ground state (for negative $\epsilon$) into the (0,1) excited
state, or vice versa.

\begin{figure}[t]
\vspace{1.5cm}
\begin{center}
\includegraphics[width=8.6cm]{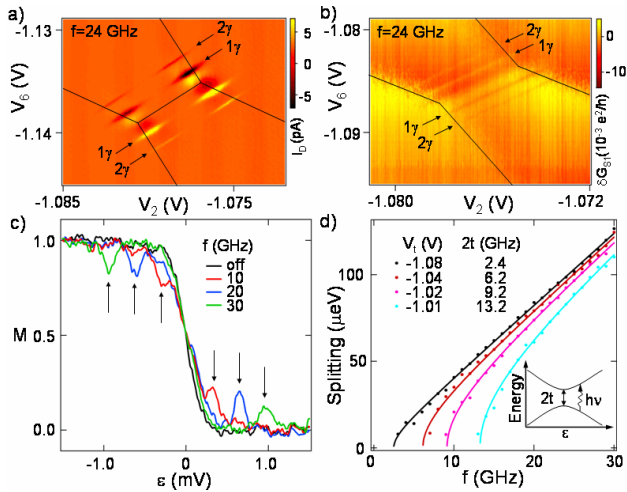}
\end{center}
\vspace{-2.5cm} \caption{(a) $I_D$ as a function of $V_{2}$ and
$V_{6}$ with $V_{sd}$=0 $\mu$V. 24 GHz photons are applied to gate
4. Single and two photon processes are visible. Charge transitions
are marked with black lines. (b) $\delta G_{S1}$ as a function of
$V_{2}$ and $V_{6}$ with 24 GHz photons applied to gate 4. A
best-fit plane has been subtracted from the data. (c) Number of
electrons on the upper dot, $M$, as a function of $\epsilon$ for
several microwave frequencies. (d) One-half of the resonance peak
splitting as a function of $f$ for several values of $V_t$. Solid
lines are best fit theory curves. Inset: two-level system energy
level diagram. The interdot tunnel coupling $t$ results in a
splitting of 2$t$ at $\epsilon$=0.} \vspace{-0.5cm}
\end{figure}

Measurements of the frequency dependence of the resonance confirm
that these features are due to a microwave induced repopulation of
charge states. The black curve in Fig.\ 3(c) shows the measured
charge on the upper dot, $M$, as a function of $\epsilon$, in the
absence of microwave excitation. Application of microwaves to gate
4 results in resonant peaks in $M$ vs.\ $\epsilon$ that move to
larger $|\epsilon|$ with increasing frequency. A detailed
measurement of the resonant peak position as a function of
microwave frequency, $f$, is used to extract $t$ for various $V_t$
(see Fig.\ 3(d)) \cite{Oosterkamp_Nature_1998}. The peak positions
depend linearly on $f$ at high frequency. At low $f$, the interdot
tunnel coupling modifies the linear dependence. The x-axis
intercept gives the value of $2t$. For each value of $V_t$, the
experimental data have been fit using $\alpha \epsilon$=
$\sqrt{(hf)^2-(2t)^2}$, where $\alpha$ is the lever arm. $\alpha$
and $t$ were free parameters for each curve, and $\alpha$ only
changes by $\sim$20$\%$ over the range of $V_t$ used in Fig.\ 3.
The best fit for $V_t$=-1.08 V is $\alpha$=0.13$\pm$0.02,
consistent with the lever arms obtained by measuring the QPC
response as a function of $\epsilon$ at higher temperatures
\cite{DiCarlo_PRL_2004}. The data in Fig.\ 3(d) have been
converted to energy using the best fit values of $\alpha$. Our
experimental data are well fit by theory and show a variation in
$t$ of roughly a factor of 6. In addition, the $t$ values in Fig.\
3(d) agree to within 25\% with the $t$ values obtained from the
data in Fig.\ 2 using Eq.\ 1 \cite{DiCarlo_PRL_2004}. The slight
discrepancy in the 2$t$ values for $V_t$=-1.01 V is due to error
in extracting the lever arm for the data in Fig.\ 2 from
temperature scans \cite{noise}.

The resonant response of a two-level system can be used to extract
information about the charge relaxation and decoherence times, as
used, for instance, in the recent analysis of the Cooper pair box
\cite{Lehnert_PRL_2003}. Measurements of the resonance peak height
as a function of time after the system is moved out of resonance
and measurements of the peak width can be used to determine the
charge relaxation time $T_1$ and the inhomogeneous charge
decoherence time $T_{2}^{*}$ \cite{Abragam}.

\begin{figure}[t]
\vspace{1.5cm}
\begin{center}
\includegraphics[width=8.6cm]{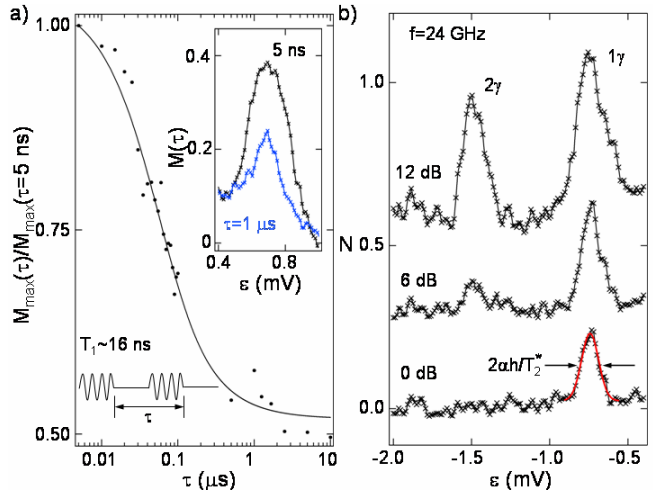}
\end{center}
\vspace{-2.5cm} \caption{(a) Resonance amplitude, expressed as
$M_{max}$($\tau$)/$M_{max}$($\tau$=5 ns), as a function of chopped
cw period, $\tau$, with $f$=19 GHz. Theory gives a best fit
$T_1$=16 ns (solid line, see text). Inset: 1$\gamma$ peak shown in
a plot of $M$ as a function of $\epsilon$ for $\tau$=5 ns and 1
$\mu$s. (b) Power dependence of the resonance for $f$=24 GHz.
Multiple photon processes occur at higher powers. Curves are
offset by 0.3 for clarity.} \vspace{0 cm}
\end{figure}

To measure $T_1$ the resonance peak height is measured as
microwaves are chopped at varying periods, $\tau$, with a 50\%
duty cycle using a fast mixer circuit \cite{Marki}. We model the
system response with a saturated signal while microwaves are
present, followed by an exponential decay with a characteristic
time scale $T_1$ when the microwaves are turned off. Taking the
time average, we expect:
\begin{equation}
\frac{M_{max}(\tau)}{M_{max}(0)}=\frac{1}{2}+\frac{T_1(1-e^{-\tau/(2
T_1)})}{\tau}
\end{equation}
For $\tau$$\gg$$T_1$, the exponential tail due to the finite
relaxation time represents an insignificant part of the duty
cycle, and the QPC detectors measure the time average of the
on/off signal, giving a resonant feature with half the height
found in the limit $\tau$$\rightarrow$0. For very short periods,
such that $\tau$$\ll$$T_1$, the charge has little time to relax,
and the QPC response is close to saturation (saturation is defined
as $M_{max}$=0.5 on resonance). When $\tau$$\sim$$T_1$, the QPC
signal is strongly dependent on $\tau$. To avoid artifacts due to
the finite rise time of the mixer circuit, we present data for
$\tau\geq$5 ns. In Fig. 4(a), we plot
$M_{max}$($\tau$)/$M_{max}$($\tau$=5 ns) as a function of $\tau$.
The experimental response is in good agreement with this theory,
and gives a best fit $T_1$=16 ns.

A measure of the charge decoherence time can be extracted from the
resonance peak width. The charge sensor response is an ensemble
measurement (in time), so we associate the peak width with the
inhomogenous decoherence time, $T_{2}^{*}$
\cite{Lehnert_PRL_2003,Abragam}. In Fig.\ 4(b) we plot $N$ as a
function of $\epsilon$ for increasing microwave powers. At low
power, only the 1$\gamma$ peak is visible. As the power is
increased the 1$\gamma$ peak approaches saturation and a 2$\gamma$
peak develops \cite{population_inversion}. A gaussian fit to the
low power 1$\gamma$ peak is shown in red in Fig.\ 4(b). We find a
half-width of 0.077 mV, which corresponds to an energy of 10.2
$\mu$eV when taking into account the lever arm. Converting this
into a time results in a lower bound $T_{2}^{*}$=400 ps. The
measurement of $T_{2}^{*}$ is sensitive to charge fluctuations,
which will broaden the resonant feature, resulting in a smaller
value for the decoherence time. Thus our measured $T_{2}^{*}$ is a
worst-case estimate.

We can compare the results of our $T_{1}$ and $T_{2}^{*}$ analysis
with other recent experiments
\cite{Hayashi_PRL_2003,Fujisawa_Nature_2002}. Fujisawa \textit{et
al.} \cite{Fujisawa_Nature_2002} have measured the energy
relaxation time in a vertical quantum dot using a pulsed gate
experiment. From a measurement of the transient current as a
function of pulse time they extract $T_1$=10 ns, which is limited
by spontaneous emission of a phonon. The $T_{2}$ time of a charge
state in a many electron ($N$$\sim$25) double dot has recently
been measured by Hayashi \textit{et al.}\ \cite{Hayashi_PRL_2003}.
From the envelope of the decay of Rabi oscillations as a function
of time Hayashi \textit{et al.} extract a $T_{2}$ time on the
order of 1 ns, which serves as an upper bound estimate for
$T_{2}^*$. The $T_{1}$ and $T_{2}^{*}$ values that we obtain from
charge sensing are in good agreement with the results of these
previous experiments.

In conclusion, we have used QPC charge sensing to detect microwave
manipulation of a single electron in a double quantum dot.
Analysis of the resonance position as a function of frequency
allows us to extract the interdot tunnel coupling. Since this
method does not rely on transport, it may be used to determine the
interdot tunnel coupling when the double dot is completely
isolated from the leads. Time-domain experiments and measured
line-widths allow us to extract $T_{1}$=16 ns and $T_{2}^{*}$=400
ps for the charge two-level system. In addition, we have
demonstrated tunability of $t$ in the few-electron regime, which
is a crucial element for many spin manipulation proposals based on
fast control of the exchange interaction.

\begin{acknowledgments}
We acknowledge useful discussions with Amir Yacoby, Leo DiCarlo,
Jacob Taylor, and Mikhail Lukin. Dominik Zumb\"{u}hl, James
Williams, and Abram Falk provided experimental assistance. We
thank M. H. Devoret, R. J. Schoelkopf, and their research groups
for very useful advice concerning microwave techniques. This work
was supported by the ARO under DAAD55-98-1-0270 and
DAAD19-02-1-0070, DARPA under the QuIST program, the NSF under
DMR-0072777 and the Harvard NSEC.
\end{acknowledgments}


\begin{thebibliography}{99}
\bibitem{noise}The lever arms for the data in Fig.\ 2 are
determined by increasing temperature until the charge transition
is thermally broadened. Increased charge noise at higher
temperatures resulted in $\sim$20$\%$ error in the lever arm for
the $V_t$=-1.01 data in Fig.\ 2.
\end{thebibliography}
\end{document}